\documentclass{epl}
\usepackage{epsfig}

\newcommand{\equ}[1]{(\protect\ref{#1})}
\newcommand{\si}{\sigma}
\newcommand{\pt}[1]{\frac{\partial {#1}}{\partial t}}
\newcommand{\Uo}{\Upsilon_0}

\newcommand{\G}[1]{G(#1, t)}
\newcommand{\Gs}[1]{G_\si(#1, t)}

\title{Breaking of scale-invariance symmetry in adsorption
  processes}
\shorttitle{Scale-invariance breaking in adsorption}
\author{R. Pastor-Satorras\inst{1} and J. M. Rub{\'\i}\inst{2}}
\institute{
  \inst{1} The Abdus Salam International Centre for Theoretical
  Physics (ICTP), Condensed Matter Section,
  P.O. Box 586, 34100 Trieste, Italy\\
  \inst{2} Departament de F{\'\i}sica Fonamental, Facultat de F{\'\i}sica,
  Universitat de Barcelona
  Diagonal 647, 08028 Barcelona, Spain
}

\begin{document} 

\maketitle

\begin{abstract}
  Standard models of sequential adsorption are implicitly formulated
  in a {\em scale invariant} form, by assuming adsorption on an
  infinite surface, with no characteristic length scales. In real
  situations, however, involving complex surfaces, intrinsic length
  scales may be relevant. We present an analytic model of continuous
  random sequential adsorption, in which the scale invariance symmetry
  is explicitly broken. The characteristic length is imposed by a set
  of scattered obstacles, previously adsorbed onto the surface. We
  show, by means of analytic solutions and numerical simulations, the
  profound effects of the symmetry breaking on both the jamming limit
  and the correlation function of the adsorbed layer.
\end{abstract}

The irreversible adsorption of colloidal particles onto a solid
surface \cite{bartelt91,evans93} has been the subject of a
particularly active research effort over the last years. The kinetics
of adsorption has been mainly studied through the formulation of
different models, aiming to capture the essential features of the
process. These models are defined via a set of rules by which the
particles accommodate when arriving at the surface, and their main
purpose is to reproduce the experimentally observed properties of the
adsorbed phase. Among those, we emphasize the maximum density of
adsorbed particles ---the {\em jamming limit} $\rho_\infty$--- and the
structure of the adsorbed layer, as measured by the correlation
function $g(x)$ \cite{evans93}. In the basic models proposed so far
(the random sequential adsorption model (RSA)
\cite{flory39,renyi58,feder80,schaaf88,senger91,ramsden93}, the
ballistic model \cite{meakin87,talbot92,thompson92}, and their
subsequent extensions), the kinetics is considerably simplified, both
numerically and analytically, by assuming that the adsorption takes
place onto a {\em planar} surface of {\em infinite} extension.  Under
this condition, and given that there is no characteristic length
involved in the problem (apart from the size of the particles) it is
presumable that the basic dynamic quantities satisfy certain scaling
laws. The existence of those scaling laws is ultimately responsible
for the fact that, once the model is established by predicating a
certain set of rules, all relevant quantities remain fixed and their
values depend only on the rules adopted. In particular, all properties
are independent of the particle size and the rate of adsorption---this
latter assumed to be constant in space and time.  Moreover, in the
absence of external forces or interactions among the particles
\cite{pago95,adamczyk96,pastor98a}, the adsorbed layer exhibits a
simple structure in which correlations decay very fast, thus
indicating absence of long range order.  Although models formulated
under those conditions may accurately reproduce many experimental
situations, their extension to more complex surfaces having an
intrinsic structure is by no means trivial. For example, a
characteristic length may be present in the substrate, which may alter
the kinetics of the process and the structure of the adsorbed layer.
With the only exception of some discrete models formulated to analyze
adsorption in the presence of point-like quenched impurities
\cite{milosevic93,lee96}, and a model of RSA of spherical particles
adsorbing onto a substrate composed by a random collection of points
\cite{jin93}, the physics on those complex substrates (beyond the
scaling regime) remains essentially unexplored.

Our purpose in this Letter is to show that when the scale invariance
is broken by introducing a characteristic length, new aspects of the
problem arise, leading to considerable changes in the adsorbed layer.
The appearance of new scales may originate from the existence of
pinned objects on the surface \cite{milosevic93,lee96}. Thus, to make
our analysis concrete, we will present an extensive study of a
one-dimensional (1d) model in which particles adsorb according to the
RSA rules onto a line where a distribution of spherical obstacles of
different size has been previously dispersed.  We note that this model
is essentially different from those of
Refs.\cite{talbot89,meakin92,senger93,adamczyk97,pastor99}, which deal
with the {\em simultaneous} adsorption of particles of different
sizes.

In the RSA model \cite{evans93,renyi58}, the adsorbing particles are
sequentially located at random positions on the surface. If an
incoming particle overlaps with a previously adsorbed one, it is
rejected; otherwise, it becomes irreversibly adsorbed. The continuous
version of this model can be solved in 1d by analyzing the density
function of gaps---holes between two consecutively adsorbed particles.
In the RSA of spheres of a single diameter $\si$, the density of gaps
of length $x$, $\Gs{x}$, fulfills the equations \cite{talbot92}
\begin{eqnarray}
  \pt{\Gs{x}} & = & -(x-\si) \Gs{x}  
  +  2 \int_{x+\si}^\infty \Gs{y} d y, \:\:\:  x\geq\si;  \label{eq:single1} \\ 
    \pt{\Gs{x}} & = &  2 \int_{x+\si}^\infty \Gs{y} d  y, \:\:\:  x\leq\si.
  \label{eq:single2} 
\end{eqnarray}
Here time $t$ has been rescaled by the rate of arrival of the
particles to the line and it has therefore units of inverse length.
The subscript $\si$ explicitly denotes the dependence on the size of
the particles.  The coverage $\rho_\si(t)$ is given by $1-\int_0^\infty x
G_\sigma(x,t) d x$. The solution of Eqs.~\equ{eq:single1} and
\equ{eq:single2}, with the initial condition of an infinite clean
substrate [i.e., $G_\si(x, 0)=0$] yields the result $\rho_\si(t)=\psi(\si
t)$ \cite{talbot92}, where $\psi (t) \equiv \int_0^t F(u) d u$ is the coverage
corresponding to an RSA process with particles of size $1$. We have
defined the usual function $F(t) \equiv \exp\left\{ -2 \int_0^t d z
  (1-e^{-z})/z \right\}$.

From this solution we see that $\rho_\infty=\lim_{t\to\infty} \rho_\sigma(t)\equiv\rho_R=0.74759$
\cite{renyi58}, independent of $\sigma$. This fact can be understood by
noticing that the gap distribution, as given by Eqs.~\equ{eq:single1}
and \equ{eq:single2}, fulfills the identity
\begin{equation}
  G_{\lambda \si}(x, t) \equiv  \lambda^{-2} G_{\si}(\lambda^{-1} x, 
  \lambda t),
\end{equation}
for any real positive number $\lambda$.  This identity means that the gap
density $G_\si$ is {\it scale invariant}: covering the line (and in
general any surface) with particles which are larger by a factor $\lambda$
has the only effect of reducing the time at which a given
configuration is reached, by a factor of $\lambda^{-1}$.  It is the
existence of this scale invariance that is responsible for the fact
that the jamming limit remains fixed upon variations of the particle
size $\sigma$.

The presence of impurities---defined as preadsorbed obstacles of a
fixed size $\si_0\neq\si$---breaks the scale invariance by introducing an
external length scale. The jamming limit and the structure of the
adsorbed phase must therefore depend in this case on the size of the
adsorbing particles, more precisely, on the size ratio of particles to
obstacles, $r=\si/\si_0$.

We consider in general the adsorption of particles of size $\si$ onto
a linear substrate, over which there is a preadsorbed set of obstacles
of size $\si_0$, present with an initial density $\rho_0$.  Assuming that
the impurities have been adsorbed onto the surface following the RSA
dynamics, they reach the coverage $\rho_0$ at time $t_0$,
\begin{equation}
  \rho_0 = \psi(\sigma_0 t_0) = \psi(\zeta_0).
  \label{eq:rho0}
\end{equation}
The dependence of $\rho_0$ on $t_0$ is through the combination $\zeta_0=\si_0
t_0$. At this time, the surface exhibits a gap distribution given by
$G_{\si_0}(x, t_0)$.  The problem translates then to solving
Eqs.~\equ{eq:single1} and \equ{eq:single2} with the initial condition
that, at time $t_0$, the gap density is given by $G_0(x) \equiv
G_{\si_0}(x, t_0)$.

We contemplate two possibilities, namely $\sigma > \sigma_0$ and $\sigma < \sigma_0$.  The
instance $\sigma > \sigma_0$ has been previously considered in the literature,
numerically \cite{milosevic93,lee96,lee97} and analytically
\cite{lee96,lee97,bennaim94}, on one- and two-dimensional lattice
(discrete) models. In this case, the imposed characteristic length is
small, thus implying slight variations of the form of the surface with
respect to the planar form. The first characteristic we want to show
is that our model admits in this case a very direct solutions, which
agrees with the continuum limit of the lattice models previously
proposed.

To show this, we perform in Eq.~\equ{eq:single1} the substitution
$\G{x}=e^{-(x-\si)t} H(t)$. The corresponding equation for $H(t)$ is
then $d H/d t =(2 e^{-x \si}/t)H$, which is solved with the initial
condition $H(t_0)=H_0$.  The constant $H_0$ is determined by
comparison with the initial value $G^>_0(x) \equiv G_0(x>\si_0)$.  We
obtain
\begin{equation}
  G(x, t) = \Uo t^2 e^{-(x-\si)t} F(\si t), \:\:\:  x>\si,
  \label{eq:smallsigmalargex}
\end{equation}
where we have defined the constant 
\begin{equation}
\Uo = e^{-(\si-\si_0)t_0}\frac{F(\si_0 t_0)}{F(\si t_0)}  \equiv
  e^{-(r-1) \zeta_0} \frac{F(\zeta_0)}{F(r \zeta_0)}.
\end{equation}
Eq.~\equ{eq:single2} is solved by direct integration, yielding
\begin{equation}
  G(x, t) = G_0(x) + \Uo \int_{t_0}^t 2 u  e^{-x u}
  F(\si u) d  u, \:\:\:  x<\si,
\end{equation}
where $G_0(x)$ equals $G^<_0(x) \equiv G_0(x<\si_0)$ or $G^>_0(x)$,
according to the value of $x$.  Further integration of the quantity
$xG$, in the limit $t\to\infty$, yields the jamming limit
\begin{equation}
  \label{eq:coverbigger}
  \rho_\infty = \rho_0 + \Uo \left[ \rho_R - \psi(r \zeta_0)
  \right].
\end{equation}
This result coincides with the continuum limit obtained from the
lattice model in Ref.~\cite{lee97}. 

We turn now to the more complex case $\si<\si_0$, in which more
interesting changes are expected, due to the fact that the planar form
of the surface is considerably altered.  The difficulty in solving the
model stems from the coupling of the solution to the initial
conditions. We consider in particular the case $\sigma_0/2 \leq \sigma \leq \sigma_0$; the
case $\si<\si_0/2$ can be worked out along the same lines.  The
solution of the model is found for different ranges of values of $x$:

($a$) $x>\si_0$: From Eqs.~\equ{eq:single2} and
\equ{eq:smallsigmalargex}, we readily obtain
\begin{equation}
  G^{(a)}(x,t) = \Uo t^2 e^{-(x-\si)t} F(\si t).
\end{equation}

($b$) $\si_0>x>\si$: From Eq.~\equ{eq:single1}, we see that the
equation is coupled to $G^{(a)}$. The solution is
\begin{equation}
  G^{(b)}(x,t) = e^{-(x-\si)(t-t_0)}  \int_0^{t_0} 2 u e^{-u x}
  F(\si_0 u) d u 
  + \Uo e^{-(x-\si)t} \int_{t_0}^t 2 u e^{-\si u} F(\si u) d  u.
\end{equation}

($c$) $\si>x>\si_0-\si$: Given $G^{(a)}$, the solution follows by
direct integration:
\begin{equation}
  G^{(c)}(x,t) = G^<_0(x)  + \Uo \int_{t_0}^t 2 u e^{-x u} F(\si u) d  u.
\end{equation}

($d$) $\si_0-\si>x>0$: The equation for $G^{(d)}$ is coupled to the
cases $a$ and $b$.  The solution is found to be
\begin{eqnarray}
  G^{(d)}(x,t) &=& G^<_0(x) 
  + 2 \int_{t_0}^t d  v \int_0^{t_0} 2 u
    e^{-\si u}{\cal L}_x (u+v-t_0) F(\si_0 u) d  u \nonumber \\
  &+& 2 \Uo \int_{t_0}^t d  v {\cal L}_x (v)   \int_{t_0}^v 2 u
    e^{-\si u} F(\si u) d  u 
    + \Uo \int_{t_0}^t 2 u e^{-(\si_0-\si)u} F(\si u) d u,
\end{eqnarray}
where we have used the auxiliary function
\begin{equation}
  {\cal L}_x (z) = \frac{1}{z} \left\{e^{-x z} - e^{-(\si_0-\si)z}\right\}
\end{equation}

In order to estimate the covering at jamming, we compute the integral
$\int_0^\infty x G(x,t) dx$. In the limit $t\to\infty$ the contribution from regions
$a$ and $b$ vanishes, since at jamming there are no gaps of length
larger than a particle diameter.  After performing some cumbersome
algebra, we arrive at the final expression for the total coverage,
counting both particles and obstacles:
\begin{eqnarray}
  \rho_\infty &=&  \rho_0 (1+r) + 2 r \int_0^{\zeta_0} ( 
    e^{-r u} - 2 e^{-u}) F(u) d  u \nonumber \\
    &+&  2 r \Uo  \int_{\zeta_0}^\infty \left\{
    \left[ 1 + (1-r)u \right] e^{-(1-r)  u} 
    +  e^{-r u} -  2 e^{-u} \right\} F(r u) d  u. 
  \label{eq:coversmaller}
\end{eqnarray}

\begin{figure}[t]

  \centerline{\epsfig{file=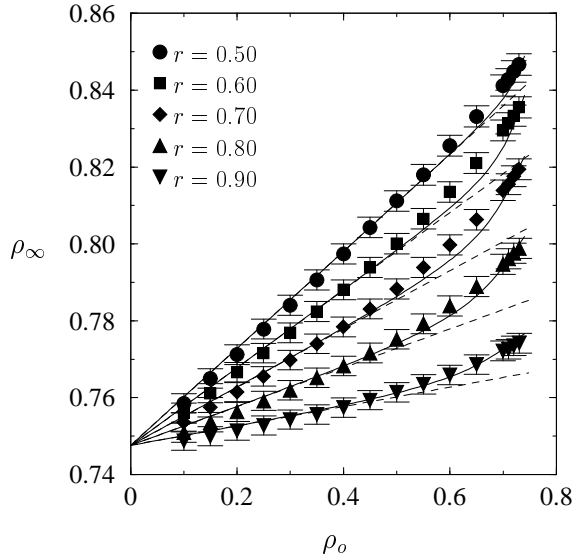, width=7.5cm}}

  \caption{Jamming limit $\rho_\infty$ as a function of the initial
    density $\rho_0$. Full lines, analytic result
    Eq.~\equ{eq:coversmaller}; filled symbols, computer simulations;
    dashed lines, linear approximation Eq.~\equ{eq:taylor}.}
  \label{fig:jammings}
\end{figure}

Equation \equ{eq:coversmaller} determines the coverage at jamming for any
value of $\rho_0<\rho_R$, both explicitly and implicitly as a function of
the initial time $t_0$. We can however determine an explicit form as a
function of $\rho_0$ in the limiting case $\zeta_0 \ll 1$ (small initial
coverage $\rho_0$), by Taylor expanding the expression for $F(t)$. We
obtain
\begin{equation}
  \rho_\infty = \rho_R + (1-r) (1-\rho_R) \rho_0  + {\cal O}(\rho_0^3).
\label{eq:taylor}
\end{equation}
That is, up to corrections of order $\rho_0^3$, $\rho_\infty$ grows linearly with
$\rho_0$, with slope $(1-r) (1-\rho_R)$. Incidentally, we note that the same
Taylor expansion in valid for the expression \equ{eq:coverbigger},
corresponding to $\si>\si_0$. In this case, however, since $r>1$, the
jamming limit decreases linearly with $\rho_0$.

For larger values of $\rho_0$ we can estimate the theoretical predictions
of Eq.~\equ{eq:coversmaller} by integrating this expression up to a
very large time $t$. Given a value of $\rho_0$, the corresponding time
$t_0$ is found by numerically solving Eq.~\equ{eq:rho0}. Having
obtained these two values, we perform the integration in
Eq.~\equ{eq:coversmaller}.  Figure \ref{fig:jammings} shows the
results of such integration for different values of $r$, as a function
of the initial density $\rho_0$.  The symbols represent data obtained
from Monte Carlo simulations of the model on a line of length
$L=5000\si$ with periodic boundary conditions, averaging over 100
realizations. The error bars reported are standard deviations. The
dashed lines correspond to the results obtained in the linear
approximation \equ{eq:taylor}.

From Figure \ref{fig:jammings} we observe that the jamming limit is
very well approximated by the linear expression \equ{eq:taylor}, up to
a critical density $\rho_0^{(c)}$, above which $\rho_\infty$ overshoots and
increases faster. The critical density can be seen to depend
approximately linearly on the size ratio, being a decreasing function
of $r$. The presence of this critical density can be understood as
follows: The effect of a small concentration of obstacles is to
essentially impose a small perturbation in the structure of the
adsorbed phase. For $r\ll1$ and $\rho_0\ll1$, one can consider the
interactions of particles and impurities as effectively decoupled. We
are then in a situation where the particles saturate the surface up to
a density $\rho_R$, leaving free a fraction $1-\rho_R$ which is filled with
impurities with density $\rho_0$.  This amounts effectively to a total
coverage $\rho_\infty \approx \rho_R + (1-\rho_R)\rho_0$, which indeed coincides with the
exact Taylor expansion, Eq.~\equ{eq:taylor}, in the limit $r\to0$. We
can thus interpret this case as a {\em soft} symmetry breaking, which
induces at most linear corrections to the jamming limit.  When the
density $\rho_0$ increases, the interaction between particles and
impurities grows larger. Beyond $\rho_0^{(c)}$, the presence of
impurities radically alters the structure of the adsorbed phase,
breaking completely the scale invariance symmetry, and inducing
non-linear corrections to the jamming limit.

\begin{figure}[t]
  \centerline{\epsfig{file=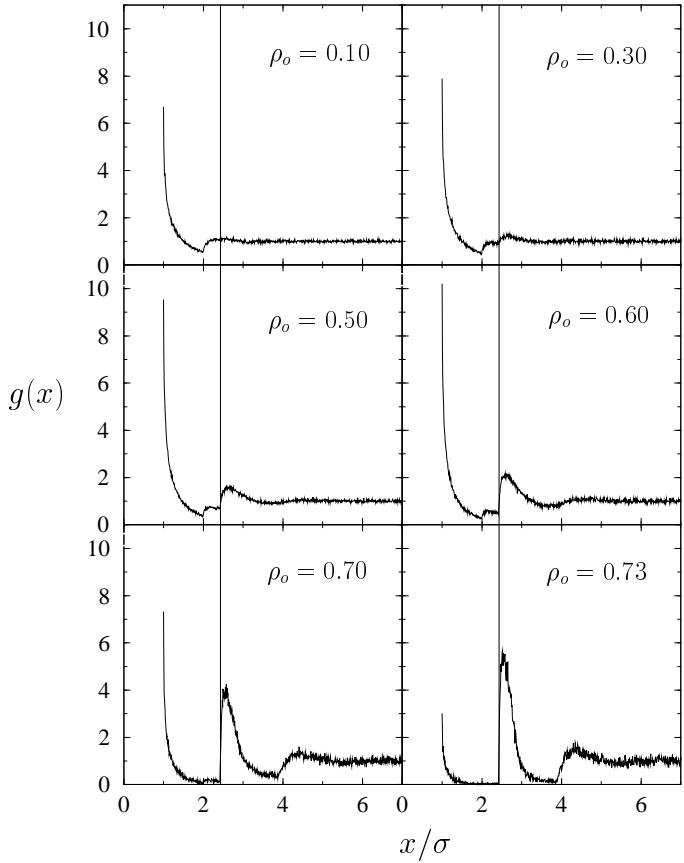, width=9cm}}
  \caption{Particle-particle correlation function for fixed $r=0.70$
    and different initial coverages $\rho_0$.}
  \label{fig:correls}
\end{figure}

We can assess the effects of the symmetry breaking on the structure of
the adsorbed phase by studying the particle-particle correlation
function $g(x)$. In Figure \ref{fig:correls} we have plotted $g(x)$ as
a function of the reduced length $x/\sigma$, for different values of $\rho_0$
and a fixed size ratio $r=0.70$. Data is obtained from simulations
onto a line of length $L=10000\sigma$, averaging over 10 different
realizations. The bin width used is $1/100$.  For this particular
value of $r$, we can estimate from Fig.~\ref{fig:jammings}
$\rho_0^{(c)}\approx0.50$. For values of $\rho_0<0.50$, we observe that the
correlation function has essentially the same shape as in standard
RSA. At $\rho_0=0.50$, however, we observe the development of a secondary
peak in $g(x)$, which eventually grows and takes over for larger
concentrations. This peak corresponds to a majority of pairs of
particles separated by exactly one obstacle, such that the distance
between the centers of the particles in a pair is $x=\sigma+\sigma_0$ or $x/\sigma
=1+1/r\simeq2.429$. This value has been marked in Fig.~\ref{fig:jammings}
by means of a straight vertical line. For even larger values of the
initial concentration, $\rho_0\geq0.70$, a third peak in the correlations is
also observed.

In summary, we have studied the kinetics of adsorption onto a line
under the conditions for which the density of gaps among particles is
not scale invariant. The analysis of the process, governed by the RSA
rules, has revealed the existence of new and rich phenomenology beyond
the scaling regime, including: substantial increase in the jamming
limit, which depends on the density of inhomogeneities introduced; the
appearance of new correlations among the particles; and also the
existence of a critical density of impurities, dividing regimes of
soft and complete violation of the scaling symmetry. To this purpose,
we have proposed and solved analytically a continuous 1d model in
which particles adsorb, following the RSA rules, onto a surface in
which spherical particles of different size have been previously
scattered.  When particularized to the case in which the imposed
length scale is smaller than the size of the particles, our model
reproduces the results obtained from a previous model considering the
adsorption on a lattice in the presence of point-like impurities. Our
results may offer new perspectives on what concerns modelization of
the adsorption phenomena in more complex surfaces, as the ones having
a certain roughness or an intrinsic structure manifested through the
existence of a characteristic correlation length. In those surfaces,
found in situations of practical interest, the property of
scale-invariance, inherent to standard adsorption models, is no longer
satisfied.

\vspace*{0.25cm}
\centerline{***}
\vspace*{0.25cm}

We thank M. A. MU{\~N}OZ for helpful discussions.  The work of RPS has
been supported by the European Network under Contract No.
ERBFMRXCT980183. JMR acknowledges financial support by CICyT
(Spain), Grant No. PB98-1258.

\end{document}